\documentclass[aps,pre,twocolumn,showpacs]{revtex4}

\usepackage{bm}
\usepackage{epsfig}

\begin{document}

%\draft
%\tighten

\title
{Modulated structures in electroconvection in nematic liquid crystals}
\author{S. Komineas$^1$, H. Zhao$^2$, and L. Kramer$^1$}
\affiliation{$^1$Physikalisches Institut, Universit\"at Bayreuth, D-95440
Bayreuth, Germany \\
$^2$ Max-Planck-Institut f\"ur Physik komplexer Systeme,
N\"othnitzer Str. 38, D-01187 Dresden, Germany
} 
\date{\today}

\begin{abstract}
Motivated by experiments in electroconvection in nematic liquid crystals
with  homeotropic alignment we study the coupled amplitude equations
describing the formation of a stationary roll pattern in the presence of
a weakly-damped mode that breaks isotropy.
The equations can be generalized to describe the planarly aligned case
if the orienting effect of the boundaries is small, which can be
achieved by a destabilizing magnetic field.
The slow mode represents the in-plane director at the center of the cell.
The simplest uniform states are normal rolls which may undergo a
pitchfork bifurcation to abnormal rolls with a misaligned in-plane
director.
We present a new class of defect-free solutions with spatial modulations
perpendicular to the rolls.
In a parameter range where the zig-zag instability is not relevant these
solutions are stable attractors, as observed in experiments.
We also present two-dimensionally modulated states with and without
defects which result from the destabilization of the one-dimensionally
modulated structures.
Finally, for no (or very small) damping, and away from the rotationally
symmetric case, we find static chevrons made up of a periodic
arrangement of defect chains (or bands of defects) separating
homogeneous regions of oblique rolls with very small amplitude.
These states may provide a model for a class of poorly understood
stationary structures observed in various highly-conducting materials
(``prechevrons'' or ``broad domains'').
\end{abstract}


\maketitle

\section{Introduction}     \label{intro}

Nematic liquid crystals, the simplest type of intrinsically
anisotropic fluids, continue to provide model systems for a wide variety
of interesting nonlinear dynamical  phenomena like optical instabilities
\cite{optic}, flow-induced nonlinear waves \cite{BoeBu2000},
critical properties of nonequilibrium
transitions  \cite{conv}, and in particular electrically or thermally driven
convection instabilities \cite{conv,KrPe2000}
(see also \cite{LCbook} and references therein).

In nematics the mean orientation of the rod-like molecules is described
by the director $\hat n$.
Electroconvection (EC) driven by an ac voltage $U$ at frequency $\omega$
is commonly  observed  in thin nematic layers
sandwiched between glass plates with transparent electrodes
using nematics with  positive conductivity anisotropy ($\sigma_a>0$) and  
negative or slightly positive dielectric anisotropy $\epsilon_a$.
In the well studied {\it planarly} aligned case, where $\hat n$ is
anchored parallel
to the bounding plates along a direction which we will take as $x$ (we choose
the layer in the $x$-$y$ plane)
EC sets in directly from the homogeneous state at a critical voltage
$U_c(\omega)$ and leads slightly above threshold to ordered roll patterns
associated with a periodic director distortion with the critical wave
vector ${\vec q_c}(\omega)$. 
Here we will only consider the most common situation where the bifurcation is
supercritical and leads to stationary rolls  with ${\vec q_c}$ parallel
to ${\hat x}$ (normal rolls (NRs)). 
In the usual low-frequency conduction regime, where the wavelength is
controlled by the cell thickness, this may exclude in particular very
low frequencies where the rolls at 
threshold may be oriented obliquely (depending on the material). 
In NRs near threshold the director remains in the $x$-$z$ plane,
i.e., perpendicular to the roll axis.

The investigation of homeotropically oriented cells using nematics
with manifestly negative dielectric anisotropy $\epsilon_a<0$ was
initiated rather recently,
see \cite{richter94,richter95a,richter95b,kai97} for
experimental and \cite{hertrich,RoHe96,KraPe95} for theoretical work.
In this case the director is initially  oriented perpendicular to the layer, 
i.e., in the $z$ direction,
so the system is isotropic in the $x$-$y$ plane.
Then the first instability is the spatially homogeneous
Freedericksz transition where the director bends away from the
$z$ direction, singling out spontaneously a direction $\hat{c}$ in the
$x$-$y$ plane. After the transition the slow, undamped variation of the in-plane
director $\hat{c}$ (the Goldstone mode) may be described by an angle $\phi$.
At higher voltages there is a further transition to EC
which is in many respects similar to that in cells with planarly aligned
nematics.
However, now the Goldstone mode has to be included in the description
even right at threshold. It turns out that the torque arising
when the in-plane director and the wavevector are (slightly) misaligned
is destabilizing, i.e., it acts to increase the 
misalignment (``abnormal torque''). 
Then the in-plane director in NRs is not perpendicular to the (local) roll axis.
These NRs with a misaligned in-plane director were termed ``abnormal rolls''
(ARs) \cite{richter94}.
Furthermore one may expect spatio-temporal disorder right at threshold and this
has indeed been observed, at least in the oblique roll regime, where
one expects faster dynamics \cite{RoHe96,kai97,ToBu98}.
For NRs the experimental situation is not totally clear \cite{ToBu98}.

By applying an in-plane magnetic field, which now defines the $x$ direction
and exerts an aligning torque on $\hat{c}$, the disorder at threshold
can be suppressed.
Indeed the situation then is similar to that in the planar case.
However, for a small field a small abnormal torque will suffice to overcome the
magnetic alignment, and then one has a transition to ordered ARs,
where the in-plane director is homogeneously rotated out of the $x$ direction.
For NRs this symmetry breaking is spontaneous, either to the left
or to the right, and the transition is described by a supercritical
pitchfork bifurcation, which has been
verified experimentally \cite{BuTo00,RoEb00}.

In the planarly aligned case one also has a transition to ordered ARs,
although this occurs at a distance from threshold such that a
quantitative description is more difficult.  Since now the director
is aligned at the cell boundaries the distortion of the in-plane 
director is confined to the center part of the cell leading to
a twist distortion. Incidentally, this is the reason why the 
phenomenon has been identified only recently in planar convection
\cite{PlaDe97}.
By applying a magnetic field in the $y$ direction one can now
destabilize $\hat{c}$ and thus move the AR transition downward towards
the primary instability. The AR transition merges with the primary
bifurcation when the field reaches the
strength of the twist Freedericksz field. When the two transitions are near
each other a simple reduced description can be used.  
This shows that  the two classes of systems are in many ways similar. 

Above the AR transition one often observes more complicated structures with or 
without defects.
In particular, in homeotropic EC, modulations of the AR mode have been
observed which leave the roll pattern (i.e., its phase) virtually
unchanged \cite{RoEb00,BuTo00}. Ideally, such structures can be considered
quasi one-dimensional (1D) with spatial 
variations only perpendicular to the (normally oriented) rolls. 
 We wish to address in particular such structures by studying the
simplest set of coupled amplitude equations capable of describing the
AR scenario. 
These equations were first derived for  homeotropic systems near threshold
\cite{RoHe96,axelthesis}  but with a slight generalization
they can also illustrate the planar case.
We here present for the first time 1D solutions of the appropriate type. 
We will then briefly discuss the destabilization of these 1D structures.
Finally we present a new class of fully ordered 2D solutions
occurring at higher voltage (or smaller
magnetic fields) involving periodic arrangements of defect chains
(or bands of defects).
We call them static chevrons since they are reminiscent of the dynamic
chevrons observed
in the dielectric range of EC, and more recently, also in homeotropic
convection \cite{ToBu98}.

Our results could also be of relevance for the higher-frequency dielectric
regime, where the rolls are very narrow and therefore the orienting effect
of planar boundary conditions is substantially weaker than in the
conduction range. Recently the weakly nonlinear  description of the
dielectric regime in planarly aligned systems  has been investigated
in detail  \cite{Ro00}. 

In Sec.~\ref{sec:basic} we discuss the basic equations, which take the
form of an activator-inhibitor system, and their homogeneous solutions. 
In Sec.~\ref{modulation} we study 1D modulated solutions.
Their stability  within the full 2D equations is investigated in
Sec.~\ref{stability} and in Sec.~\ref{2D} some 2D structures, in particular
defect-free solutions, are presented.
In Sec.~\ref{chevrons} we present the static chevron states.
We conclude by putting our results into a more general context, relate them to 
experiments, and present an outlook.

%%%%%%%%%%%%%%%%%%%%%%%%%%%%%
\section{Basic equations and homogeneous solutions}  
\label{sec:basic}

In situations where the in-plane director, described by an
angle $\varphi$ measured 
from the $x$ direction, becomes an active mode already near the threshold 
to NRs one can describe the system by the following set of coupled 
Ginzburg-Landau equations for the complex patterning mode $A$ and the 
slowly varying angle $\varphi$ \cite{RoHe96,axelthesis}
\begin{eqnarray}    \label{eq:A2D.unscaled}
	 \check \tau \partial_{\check t} \check A= 
    	\Big [ \varepsilon+\xi_{xx}^2 \partial_{\check x}^2+\xi_{yy}^2(\partial_{\check y}^2
     	&-& 2i q_c C_1 \varphi \partial_{\check y} - C_2 \varphi ^2)   \nonumber \\
     	-g|\check A|^2 + i \check \nu \partial_{\check y} \varphi \Big ] \  \check A\,,
	\\   \label{eq:phi2D.unscaled}
    	 \check \gamma_1 \partial_{\check t} \varphi =
    	K_1 \partial_{\check y}^2 \varphi+ K_3 \partial_{\check x}^2 \varphi  &-& T \varphi
  	\nonumber   \\
	+ {\Gamma \over 4}\,  [ -i q_c \check A^* (\partial_{\check y} 
	&-& i q_c \varphi) \check A + {\mbox{\rm c.c.}} ]\,.
\end{eqnarray}
Here we have chosen the $x$ direction along the wavevector of the NRs.
The angle $\varphi$ of the in-plane director is measured from the $x$ axis.
The validity of Eqs.~(\ref{eq:A2D.unscaled},\ref{eq:phi2D.unscaled})
is restricted to small 
values of the reduced control parameter (more precisely
$\varepsilon /g\ll 1$ and angles $|\varphi|\ll 1$). 
Special attention should be paid to the sign of the parameter $\Gamma$. 
If $\Gamma$ were positive the field $\varphi$ would be stabilized by
the roll pattern.
However, $\Gamma$ is negative, at least for the standard nematics
which have been used in relevant experiments \cite{RoHe96,axelthesis,Ro00}.
This gives rise to the transition to abnormal rolls and to interesting
dynamical phenomena. Note that $\xi_{yy}$ tends to zero at the
transition from normal to oblique rolls at threshold.

The equations can be justified most convincingly for homeotropic
orientation near the EC threshold (reduced control parameter
$\varepsilon\ll 1$). Overall rotation invariance then
requires that the three terms multiplying $\xi_{yy}^2$ in
Eq.~(\ref{eq:A2D.unscaled})
combine to $ (\partial_{\check y} - i q_c \varphi)^2$, so that $C_1=C_2=1$.
Without the isotropy-breaking term $-T \varphi $, which is realized
easily by an in-plane magnetic field (then $T=\check \chi_a   H^2$),
the angle $\varphi$
may not saturate. Then one has to resort to a globally rotational invariant
generalization of Eqs.~(\ref{eq:A2D.unscaled},\ref{eq:phi2D.unscaled}) 
\cite{axelthesis,RoHe96}. We will see that this is not always necessary.

The parameters $C_1,\ C_2$ were introduced to allow for more general situations
like in planar alignment (for structural stability one needs $C_2>C_1^2>0$).  
Then the magnetic field term in
Eq.~(\ref{eq:phi2D.unscaled}) actually models the
orienting effect of the boundaries. In the common conduction range, where the 
wavelength is of the order of the cell thickness, the orienting effects
are sufficiently strong so that interesting dynamics of $\varphi$ sets 
in at values of $\varepsilon$ which are too large to allow quantitative
description by Eqs.~(\ref{eq:A2D.unscaled},\ref{eq:phi2D.unscaled}).
Then, in particular, singular mean
flow has to be included \cite{KraDre01,berndthesis}. 
To reduce the damping of $\varphi$ one can either
apply a destabilizing in-plane magnetic field and/or go to the
dielectric range where the effect of boundaries is weaker \cite{Ro00}.
When $T$ becomes too small higher-order terms are needed in
Eq.~(\ref{eq:phi2D.unscaled}).

For calculations it is useful to introduce a scaled version of 
Eqs.~(\ref{eq:A2D.unscaled},\ref{eq:phi2D.unscaled}) for $\varepsilon>0$
\begin{eqnarray}  \label{eq:A2D.scaled}
	\tau\partial_t A = [1+\partial_x^2+
       	 \partial_{y}^2-2i c_1  \phi \partial_{ y} -  \phi ^2
	-|A|^2-i\nu\partial_y\phi] A \\
\medskip
  \label{eq:phi2D.scaled}
	\partial_t\phi = D_1 \partial_x^2\phi + D_2 \partial_y^2\phi-h\phi
	 +\left[ \frac{i}{2} A^*(c_2 \partial_y-i \phi) A+\hbox{c.c.}\right]
\end{eqnarray}
where 
\begin{eqnarray}  \label{eq:parameters}
\check A&=&(\varepsilon /g)^{1/2} A,\
\varphi=\varepsilon^{1/2}\phi/(\xi_{yy}q_c \sqrt{C_2}),\ \nonumber \\
\check x&=&\xi_{xx} \varepsilon^{-1/2} x,\
\check y=\xi_{yy}\varepsilon^{-1/2} y,\ 
\check t=2 \check \gamma_1 g  t/(\varepsilon |\Gamma| q_c^2), \ \nonumber \\
\tau&=&\check \tau |\Gamma| q_c^2/(2 \check \gamma_1g),\ 
\nu= \check \nu /(\xi_{yy}^2 q_c \sqrt{C_2}), \ \nonumber \\
D_1&=& 2 K_3 g/(|\Gamma| q_c^2 \xi_{xx}^2), \  
D_2= 2 K_1 g/(|\Gamma| q_c^2 \xi_{yy}^2), \  \nonumber \\
c_1&=&C_1/\sqrt{C_2},\ c_2=\sqrt{C_2},\ 
h= 2 T g/(|\Gamma| q_c^2 \varepsilon).
\end{eqnarray}
The damping parameter $h$ gives the ratio of the aligning
(=isotropy-breaking) torque over the abnormal torque of NRs. 
For large values of $h$ one can set $\phi=0$ and disregard 
Eq.~(\ref{eq:phi2D.scaled}). $h$ can be decreased by either
decreasing a stabilizing magnetic field (increasing a destabilizing field)
or by increasing $\varepsilon$.
Below we will show that, keeping the aligning torque fixed, one can write
	\begin{equation}  \label{h}
	h=\varepsilon_{\rm AR}/\varepsilon \ ,
	\end{equation}
where $\varepsilon_{\rm AR}$ is the reduced control parameter where
the transition from NRs to ARs takes place.
The parameter $\nu$ describes the action of the gradient of the
in-plane director on the phase of the rolls. Experimentally it can be
controlled by varying the frequency.
In this paper we will be concerned  with the range $\nu>0$ where NRs
are first destabilized by the transition to ARs (see below).
For $\nu<0$ the zig-zag instability comes in earlier.
Various features of Eqs.~(\ref{eq:A2D.scaled},\ref{eq:phi2D.scaled})
have been analyzed in \cite{RoHe96,axelthesis,hongthesis} and comparison
with experiment has given evidence for their validity.

We first discuss the homogenous solutions of
(\ref{eq:A2D.scaled},\ref{eq:phi2D.scaled})
for rolls with modulation wavevector $(Q,P)$ where
$A\!=\!A_0 e^{i(Q x+P y)},\ A_0\!\ge\!0$, and a constant in space angle
$\phi\!=\!\phi_0$.
One is left with the dynamical system
\begin{eqnarray}  \label{eq:0D}
	\tau\partial_t A_0 &=&
         [1- Q^2 -P^2 +2 c_1 P \phi_0- \phi_0^2- A_0^2] A_0\,, \nonumber \\  
	\partial_t\phi_0 &=& (A_0^2- h)\phi_0 - c_2 P A_0^2\,.
\end{eqnarray}
These equations can be classified as an activator-inhibitor system with
activator $A_0$ (positive linear growth rate for not too large wavevector)
and inhibitor $\phi_0$. 
For simplicity we will in the following consider $Q=0$
(the results are easily generalized). 
For $P=0$ (rolls exactly in the $x$ direction; we will deal with this case,
except in Sec.~\ref{chevrons}) there is the symmetry $\phi \to -\phi$.
The basic uniform state $A_0\!=\!0,\ \phi_0\!=\!0$, is an unstable solution of 
(\ref{eq:0D}), and the NR state $A_0\!=\!1,\ \phi_0\!=\!0$,
is stable only for $h\!\geq\! 1$.
Finally, the abnormal roll (AR) states 
	\begin{equation}  \label{eq:ARs}
		A_{\rm AR}=\sqrt{h},\qquad \phi_{\rm AR}=\pm\sqrt{1-h}
	\end{equation}
exist in the range $0\!\leq\! h \!<\! 1$.
They bifurcate from the NRs at $h\!=\! 1$ and they 
break the symmetry $\phi \to -\phi$ of the equations. For $h>0$ the
two AR states are global attractors with regions of attraction
$A_0 > 0,\ \phi_0 >/< 0$, respectively.
For $h>4 \tau/(1+ 4 \tau)$ the AR states represent saddles
(two real eigenvalues), otherwise spiral points.

The case $h\!=\!0$ needs special attention. Clearly the whole band
	\begin{equation}  \label{eq:hong}
		A_0 = 0, \qquad \phi_0=\mbox{constant},
	\end{equation}
is solution of the equation. The segment $|\phi_0|<1$ is repulsive
whereas the regions $|\phi_0| >1$ are attractive. The  AR state
(for $h\!=\! 0$) separates the two cases.
There is a separatrix $A_0^2+1/(1+\tau)\phi_0^2=1$, which separates
trajectories flowing out of the repulsive segment from those coming from
infinity.

The degeneracy for $h=0$ is presumably realistic for homeotropic EC.
It is a consequence of rotational invariance.
For planar EC the degeneracy is removed by higher-order terms, in particular
a term proportional to $\phi^3$ in Eq.~(\ref{eq:phi2D.scaled}). Nevertheless
it is instructive to study this limit where results simplify.

%%%%%%%%%%%%%%%%%%%%%%%%%%%%%%%%%%%%%
\section{1D modulated structures}      \label{modulation}

Next we consider modulated structures that leave the roll pattern
untouched, i.e., which do not involve the phase of the complex field $A$. 
This can occur generically only for spatial variations along $x$.
Then the equations take the form
\begin{eqnarray} 
 \tau \partial_t A &=& (\,\partial_x^2 + 1 - A^2 - \phi^2\,)\,A\,, 
     \label{eq:A} \\
 \partial_t \phi &=& (\,D_1\, \partial_x^2 + A^2 - h\,) \phi \,. \label{eq:phi}
\end{eqnarray}
In this section we will study these equations. We will choose $A>0$.

%%%%%%%%%%
\subsection{Single domain walls}

We start by discussing domain walls that connect the two variants of
AR solutions and their interaction. Near the AR transition ($h$ near 1)
domain walls attract each other, so that modulated states are unstable.
This can be seen from the fact that for $1-h\ll 1$ the amplitude can
be eliminated adiabatically from Eq.~(\ref{eq:A}) leading to 
	\begin{equation} 
	 \partial_t \phi = (\,D_1\, \partial_x^2 + 1-h- \phi^2\,) \phi \,.
                \label{eq:phi2}
	\end{equation}
In this equation all modulated states (they can be expressed in terms
of an elliptic integral) are unstable, although their lifetimes are
exceedingly long due to the exponentially 
weak (attractive) interaction between well separated domain walls (see below).

Surprisingly, for lower damping, the interaction acquires repulsive parts,
so that stable  modulated structures can emerge.
In fact, for $h \to 0$ the interaction becomes purely repulsive so
that only {\it periodic} modulations exist (see below).

Actually, for $h\!=\! 0$ a static domain wall solution can be found
analytically. It reads
	\begin{equation}   \label{eq:front}
	A_w(x) = A_0\, \hbox{sech}(\beta x), \qquad 
	\phi_w(x) = \phi_0\, \tanh(\beta x)\,,
	\end{equation}
where
	\begin{equation}  \label{coeff}
	 \beta^2 = \frac{1}{1+ 2D_1},\quad A_0^2 = 2 D_1\, \beta^2, 
                          \quad \phi_0^2 = 2(1+D_1)\,\beta^2. 
	\end{equation}
It selects  the states $A\!=\!A_0,\
\phi\!=\!\pm \phi_0$ from the continuum of states (\ref{eq:hong}). 
We have checked numerically and could find no stable static
domain wall solution other than (\ref{eq:front}). We note that
since $\phi_0^2\!>\! 1$ (\ref{eq:front}) connects stable states.

We mention that the stationary wall (\ref{eq:front}) is embedded
in a continuous family 
of moving walls connecting inequivalent states (\ref{eq:hong}). 
Their study is beyond the scopes of the present paper.

\begin{figure}
  \epsfig{file=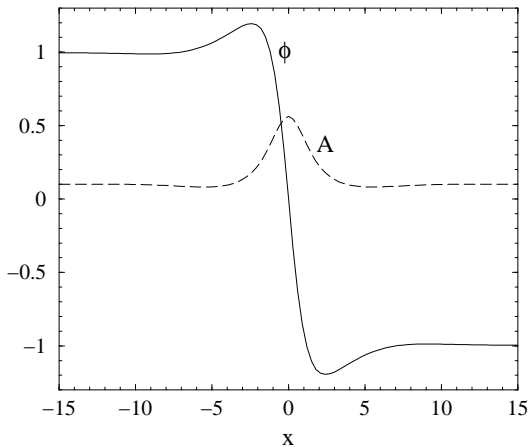,width=7.0cm}
  \caption{Domain wall solution connecting the two abnormal roll states
 $A\!=\!h,\; \phi\!=\!\pm\sqrt{1-h}$ for $D_1=0.2$ and $h=0.01$.
  }
 \label{fig:wall}
\end{figure}

Since for $h \ne 0$ the walls can connect only to the AR states and
$\phi_{\rm AR} \to \pm 1$
($\ne \phi_0 $) for $h \to 0$, this limit has to be clarified.
An example of a numerically stable (within the 1D equations) wall for small 
$h$ is shown in Fig.~\ref{fig:wall}.
The fields go to their AR values at spatial infinity. However, in the wall
region the field $\phi$ exhibits an overshoot approaching the values
$\pm\phi_0$ of (\ref{eq:front}). The overshoot becomes longer and
approaches nearer to $\pm\phi_0$ as $h\!\to\! 0$. 

In order to understand this behavior better we study the spatial decay
of the solution into the AR state. This is obtained by linearizing
Eqs.~(\ref{eq:A},\ref{eq:phi}) around an AR solution and calculating  the
spatial exponents. They are
	\begin{equation}  \label{eq:exponents}
	p_\pm^2 = h \pm \sqrt{h^2 -\frac{4 h(1-h)}{D_1}}.
	\end{equation}
For $h \to 0$ the exponents are complex and tend to zero, which explains the
slow decay. The exponents $p_\pm$ remain complex for damping constants
$h\!<\!h_{\rm osc}\!\equiv\!(1+D_1/4)^{-1}$. We will show that this is the 
bound up to which domain walls may repel and thus form
stable bound states.

%%%%%%%%%
\subsection{Interaction of domain walls, modulated structures}

In simulations of Eqs.~(\ref{eq:A},\ref{eq:phi}) one easily finds
stable stationary solutions with more than one domain wall. 
In particular there are periodic solutions which come in
two varieties: the ones that preserve the global symmetry
$\phi \!\to\! -\phi$ (``symmetric'') and others that do not preserve
it (``nonsymmetric''). 
This section is mainly  devoted to periodic solutions.
There are also pulse-type states localized around one of the AR solutions
as well as nonperiodic extended solutions. 

Let us first approach these modulated solutions from the side of
large separation  between domain walls by studying their interaction.
Repulsive interaction is a prerequisite to form stable periodic
states and interaction of both
signs are necessary to form nonperiodic arrays \cite{coullet}.
The problem is formulated as follows:
Two walls are placed symmetrically around the origin at positions $\pm x_0$.
Then the fields have zero derivative at $x\!=\!0$.
We now focus on the region $x>0$, write the amplitude and angle as 
$A\!=\!A_w(x-x_0(t)) + a(x-x_0(t),t),\;
\phi\!=\!\phi_w(x-x_0(t)) + \varphi(x-x_0(t),t)$ 
and substitute these expressions into Eqs.~(\ref{eq:A},\ref{eq:phi}). 
In the spirit of small variations {\it and} slow (``slaved'') dynamics
we keep the terms linear in $a$ and $\varphi$, but neglect the ones
that are small and contain time derivatives
	\begin{eqnarray}    \label{eq:perturbed}
	\dot{x}_0 \tau A'_w &=& -(\partial_x^2+1-3 A_w^2-\phi_w^2)\,a
               +2A_w \phi_w \varphi, \nonumber \\
	 \dot{x}_0 \phi'_w &=& -2A_w \phi_w a-(D_1\,
              \partial_x^2 + A_w -h)\, \varphi,
	\end{eqnarray}
where the dot denotes time derivatives and
the argument of all functions is $x-x_0$. 

To satisfy the boundary conditions  one needs 
$a\!=\!A_w-A_{\infty},\; a'\!=\!-A_w'$ and 
$\varphi\!=\!\phi_w-\phi_{\infty},\; \varphi' \!=\! -\phi_w'$ at $x\!=\!0$.
At the other end $x\!\to\! \infty$ the perturbations $a$ and $\varphi$
should vanish. Multiplying (\ref{eq:perturbed}) by the adjoint of the
translational mode $(A_w',-\phi_w')$ and integrating from zero
to infinity, we obtain an equation which has on the right hand side
only the boundary terms at $x\!=\!0$. On the left hand side the 
integrals may be extended to $-\infty$ with negligible error leading to
	\begin{eqnarray}  \label{eq:dotx0}
	(<\phi_w'^2> - \tau<A_w'^2>) \dot{x}_0 = A_w'^2 + A_w'' (A_w-A_{\infty})
	 \nonumber \\ 
	- D_1 (\phi_w'^2 + \phi_w'' (\phi_w-\phi_{\infty}))
	\end{eqnarray}
where $< ...> \equiv\int_{-\infty}^{\infty} dx...$ . This formula can
be used only when the
term in brackets on the left hand side is positive, i.e., for 
	\begin{equation}   \label{eq:accel}
	 \tau< \tau_{accel} \equiv <\phi_w'^2>/<A_w'^2>\,.
	\end{equation}
Otherwise translation becomes an active mode, i.e., one expects spontaneous
acceleration of the domain wall. This can be shown by retaining the small time
derivative terms omitted in Eqs.~(\ref{eq:perturbed}). We will here assume 
(\ref{eq:accel}) to hold (experimentally this appears to be the case)
and comment briefly on the acceleration instability at the end of this section.

In the special case $h\!=\!0$ and for the wall (\ref{eq:front}),
expression (\ref{eq:dotx0}) can be evaluated leading to
	\begin{equation}  \label{eq:dotx0h0}
	[ 2(1+D_1)-\tau D_1] \dot{x}_0 = 12\, D_1 \beta\, e^{-2 \beta x_0}\,.
	\end{equation}
Thus, at least for $\tau < \tau_{\rm accel}$ one has $\dot{x}_0>0$,
i.e., the interaction among walls is repulsive everywhere. 
This is consistent with the numerical observation of stable periodic solutions
and no nonperiodic states.

\begin{figure}   \epsfig{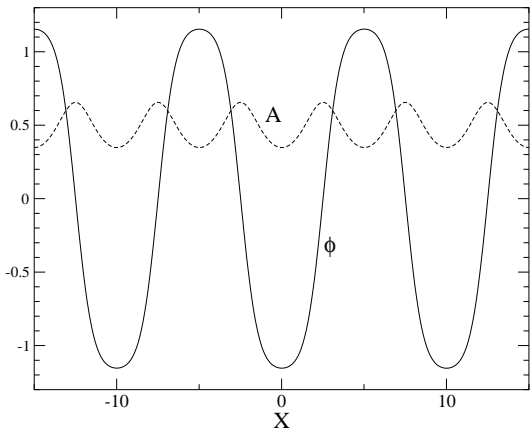}
  \caption{A symmetric periodic solution for $D_1\!=\!0.2$ and $h\!=\!0.1$.
The dashed line gives the field $A$ and the solid line the field $\phi$.
It is found by numerical simulation of
Eqs.~(\ref{eq:A},\ref{eq:phi})
with periodic boundary conditions.
  }
  \label{fig:periodic}
\end{figure}

For $h\!\neq\! 0$ we may use the asymptotic behavior characterized
by the exponents in Eq.~(\ref{eq:exponents}) to evaluate the right
hand side of (\ref{eq:dotx0}). 
In the range of complex exponents $p_\pm$, i.e., for $h\!<\!h_{\rm osc}$,
the interaction is repulsive or attractive depending on the 
distance between walls. Thus one expects stable periodic solutions
in appropriate wavelength intervals as well as nonperiodic structures,
in agreement with simulations. An example of a periodic solution is given
in Fig.~\ref{fig:periodic} for a field value $h\!=\!0.1$.

Finally, in the monotonic range $h_{\rm osc}\!<\!h\!<\!1$, the right hand side
of (\ref{eq:dotx0}) is
	\begin{equation}   \label{eq:monotonic}
	 c^2\,[\frac{D_1 p_{-}^4}{4 h(1-h)} - 1]\, \exp{(2p_{-} x_0)}\,,
	\end{equation}
where $c$ is a factor that cannot be determined from the asymptotic analysis
and $p_{-}$ is given in (\ref{eq:exponents}). One easily sees 
that the term in square brackets is negative, so that the
interaction between domain walls is now attractive.
In this range we expect neither stable periodic nor nonperiodic solutions,
although, as pointed out before, the lifetime of modulations may be very long.

\begin{figure}      \epsfig{file=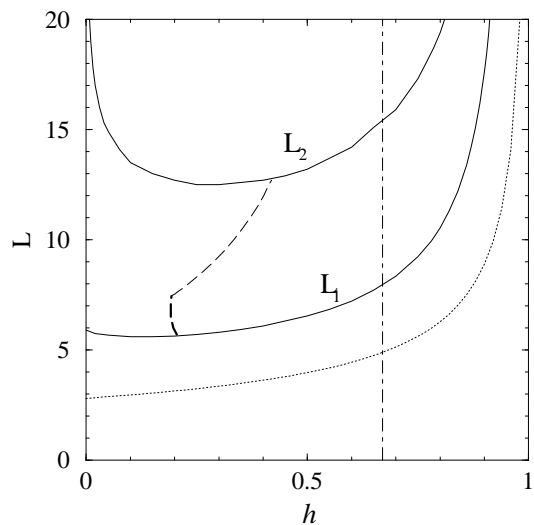,width=7.0cm}
  \caption{Stability limits of the symmetric periodic solutions in the
period (=L) vs $h$ plane for $D_1\!=\!0.2$. Below the line $L_1$ they are
unstable against breaking the $\phi \to -\phi$ symmetry, leading to ARs.
Above the line $L_2$ they are unstable against a symmetry-broken
periodic solution. The dotted line gives the minimum period
$L_{\rm min}\!=\!2\pi\sqrt{D_1/(1-h)}$ where the periodic
solutions bifurcate from NRs.
On crossing the dashed line there is a zig-zag instability which
is either short wave (thin dashed) or long wave (thick dashed).
The dash-dotted line is the limit of stability for the ARs.
  }
 \label{fig:limits}
\end{figure}

Let us now continue the symmetric periodic solutions to small wavelengths.
Since they conserve the global $\phi \to -\phi$ symmetry one expects them to
bifurcate from the NRs. This is indeed the case. From a simple linear stability 
analysis one finds that NRs are unstable with respect to periodic modes
with wavenumber $|q| <  q_c$ where 
	\begin{equation}  \label{eq:qc}
	 q_c(h)=\sqrt{\frac{1-h}{D_1}}.
	\end{equation}
Indeed, the bifurcation is of supercritical type.
In Fig.~\ref{fig:limits}, which summarizes our results on symmetric
periodic solutions,
the minimum period $L_{\rm min}=2 \pi/ q_c(h)$ is shown (dotted curve).

\begin{figure}      \epsfig{file=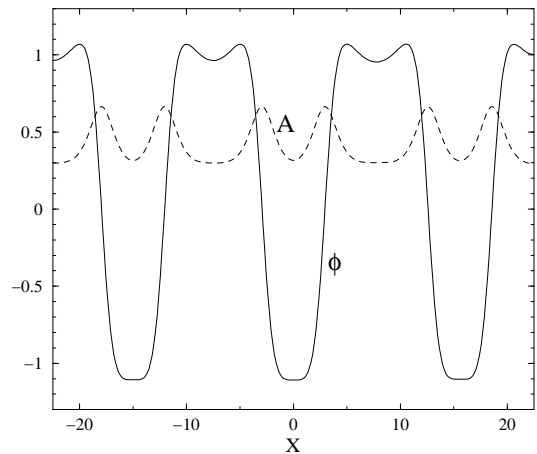,width=7.0cm}
  \caption{A non-symmetric periodic solution for $D_1\!=\!0.2$ and $h\!=\!0.1$.
The dashed line gives the field $A$ and the solid line the field $\phi$.
  }
\label{fig:nonsymmetric}
\end{figure}

When the (symmetric) periodic solutions are born they inherit the
instability of the NRs with respect to symmetry breaking.
For a fixed $h$ and by increasing the
period we find that they are stabilized at some period $L_1(h)$.
The curve $L_1(h)$ can be calculated by a linear stability analysis.
It diverges to infinity for $h\!\to\!h_{\rm osc}$
since above this value no stable periodic solutions are expected to exist
according to the asymptotic analysis given earlier. At $L_1(h)$ a branch
of unstable periodic solutions with broken $\phi \to -\phi$ symmetry
bifurcates from the symmetric periodic solutions.
We find that  the symmetric periodic solutions lose stability
(for the first time) at $L_2(h)$ in Fig.~\ref{fig:limits} 
and from here on one has a stable nonsymmetric periodic solution, 
where long and short  domains alternate.
An example of such a solution is shown in
Fig.~\ref{fig:nonsymmetric} for $h\!=\!0.1$.
Clearly this is the result of the nonmonotonic behavior of the
interaction between domain walls. Presumably the curve $L_2(h)$ also
diverges at $h_{\rm osc}$.

\begin{figure}    \epsfig{file=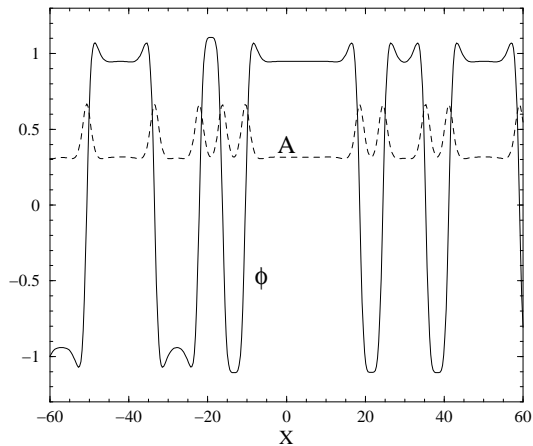,width=7.0cm}
  \vspace*{0.5cm}
  \caption{A nonperiodic solution for $D_1\!=\!0.2$ and $h\!=\!0.1$.
The dashed line gives the field $A$ and the solid line the field $\phi$.
  }
\label{fig:nonperiodic}
\end{figure}

An example of a more complicated nonperiodic solution is shown
in Fig.~\ref{fig:nonperiodic}. Such states have been also observed in
Ref. \cite{coullet}. There is no definite distance between the
walls and there seems to be no repeated structure.
The state can be spatially chaotic.

Equation~(\ref{eq:dotx0}) indicates that 
$\dot{x}_0$ diverges at $\tau\!=\!\tau_{\rm accel}$.  
For $h=0$ one finds from Eq.~(\ref{eq:dotx0h0})
	\begin{equation}  \label{eq:taccel}
	\tau_{\rm accel} = \frac{2 (1+D_1)}{D_1}.
	\end{equation}
Computer simulations show that the static wall 
is unstable, for $\tau\!\geq\!\tau_{\rm accel}$, to a steadily traveling
wall. One sees that on crossing the value $\tau_{\rm accel}$ the
wall is accelerated while the direction of motion is spontaneously chosen
\cite{hagberg97}.
No further study of moving walls will be given in this paper.
We find numerically that $\tau_{\rm accel}$, as calculated from
Eq.~(\ref{eq:accel}), increases with $h$
($\tau_{\rm accel}(h\!=\! 0) \!=\! 12,
\tau_{\rm accel}(h\!=\! 0.1)\!\simeq\!16.6,
\tau_{\rm accel}(h\!=\!0.5)\!\simeq\!26.6$ for $D_1\!=\!0.2$).
One could see this also by inspecting the form of the walls.

%%%%%%%%%%%%%%%%%%%%%%%%%%%%%%%%%%%%%%%5
\section{2D solutions}
%%%%%%%%%%%
\subsection{Stability of  1D modulations in the plane}    
\label{stability}
In view of the experimental observation of periodic modulations
it is essential to study the stability
of the solutions discussed in the previous section in the context
of the full Eqs.~(\ref{eq:A2D.scaled},\ref{eq:phi2D.scaled}) in two dimensions.
In order to reduce the number of parameters we will in this section
consider the case $c_1=c_2=1$ appropriate for homeotropic systems. 
Some results on the case when $c_1$ is smaller than $1$ will be given
in the next section.

For the AR solution (\ref{eq:ARs}) one can
go through a standard linear stability analysis which gives in the long
wavelength limit that, for $\nu\!>\!0$,
ARs become unstable against zig-zag perturbations
at $h\!=\!2/3$ \cite{RoHe96,axelthesis,hongthesis}. 
We should also recall that ARs are born from NRs as a stable
state at $h\!=\! 1$ which means that they exist stably for
	\begin{equation}  \label{eq:ARstability}
	 h_{\rm AR} \equiv \frac{2}{3} \leq h < 1\,.
	\end{equation}
The value $h_{\rm AR}$ is denoted by a dash-dotted line in 
Fig.~\ref{fig:limits}.
We use in this paper parameter values typical for the standard substance MBBA.
Specifically we take \cite{axelthesis}
	\begin{equation}  \label{eq:paramvalues}
	D_1 = 0.2, \quad D_2 = 0.5,\quad \tau = 0.5\, \quad \nu=0.6 \ .
	\end{equation}
The positive value for $\nu$ indicates that we deal with frequencies
above the codimension-2 point where the zig-zag instability of NRs is
not operative \cite{RoHe96,axelthesis,PlaDe97,RuZha98,ZhaKra00}.

In order to check the stability of the symmetric 
periodic solutions found in the previous section against two-dimensional
perturbations we have performed a numerical Floquet analysis. 
Thus we write the perturbations $a, \varphi$ of a periodic solution
$A_L(x), \phi_L(x)$ with period L as
	\begin{eqnarray}  \label{eq:perturbterms}
	a(x,y,t) = e^{\sigma t+ i s_y y} \sum_{n=-\infty}^{\infty} a_n \exp(i2\pi n x/L),
	\\
	\varphi(x,y,t) = e^{\sigma t+ i s_y y}
	  \sum_{n=-\infty}^{\infty} \varphi_n \exp(i2\pi n x/L)\,,
	\end{eqnarray}
where $a_n, \varphi_n$ are constants.
We linearize (\ref{eq:A2D.scaled},\ref{eq:phi2D.scaled}) in the perturbations
and expand the coefficient functions in the linearized equation, which have 
the periodicity L, also in a Fourier series.
After truncation a system of linear equations is obtained
which can be solved numerically to give the growth rate
$\sigma_i(s_y)$ for the linear mode with wave vector $s_y$.
The index $i$ runs over the number of Fourier modes used.
The number of modes necessary to achieve
a prescribed accuracy is found easily by numerical experimentation.

We have focused on periods which lie between the
limits $L_1$ and $L_2$ in Fig.~\ref{fig:limits}, i.e., to solutions
which are stable with respect to $x$ perturbations.
For the parameter values (\ref{eq:paramvalues})
our results are included in
Fig.~\ref{fig:limits}  (dashed line).
The thin dashes (for $7.5\!\lesssim\!L\!\lesssim\!13$) 
indicate a short wave
instability of the corresponding period when crossing the
line to the left. The thick dashes (for $5.7\!\lesssim\!L\!\lesssim\!7.5$)
indicate a long wave instability. 
It is most interesting that the present instabilities occur for a
value of $h$  considerably smaller than $h_{\rm AR}$
which means that there are stable periodic solutions for values
of $h$ for which the ARs are unstable.

We give the numerical values for
the stability limits and the wave vector $s_y$ of the unstable mode
for some periodic solutions that we
typically use in the simulations of this section (they fit into our
typical system length of 64):
\begin{eqnarray}  \label{eq:pstability}
	 \hbox{period}\ 8 : \  \  0.23 < &h& < 0.67\,, \quad s_y=0.67\,,
                \nonumber \\
	 \hbox{period}\ 9.14 : \ 0.30 < &h& < 0.74\,, \quad s_y=0.69\,,
                \nonumber\\
	 \hbox{period}\ 12.8 : \ \  0.42 < &h& < 0.85\,, \quad s_y=0.66\,. 
\end{eqnarray}
The lower limit corresponds to the short wave instability
along the $y$ direction with wave vector $s_y$.
The upper limit is due to the instability along the
$x$ direction (curve $L_1$ in Fig.~\ref{fig:limits}).

We conclude that by decreasing the parameter $h$ (which corresponds
to increasing the voltage or decreasing the magnetic field in a
relevant experiment) 
below $h_{\rm AR}$ the ARs become unstable but periodic states with
certain periods remain stable. They are then expected to be observable under
appropriate experimental conditions.
As an example we quote the experiments of Ref.~\cite{BuTo00}
for which the appropriate parameter values are close
to (\ref{eq:paramvalues}) that we use throughout this section.
It is reported that for a control parameter corresponding to
$h\!=\!0.4$ in our theory, periodic modulations of the director
orientation are observed.
Indeed this value falls into the range where ARs are unstable but some periodic states are expected to be stable.
Concerning the periodicity, the tendency towards shorter wavelengths with 
increasing voltage reported in \cite{BuTo00} is consistent with our results.

%%%%%%%%%%%
\subsection{2D modulated structures}  \label{2D}

We now proceed to investigate how the system evolves
once the periodic states are destabilized at small values of
$h$, that is, once we cross to the left of the 
dashed line in Fig.~\ref{fig:limits}.
For this purpose we have used a pseudospectral algorithm which
simulates the time evolution of Eqs.~(\ref{eq:A2D.scaled},\ref{eq:phi2D.scaled})
which we describe briefly in the following.
The linear part of the equation is tranformed into Fourier space
and the analytic formula for its time evolution is implemented in 
the algorithm. For the nonlinear part we work in real space
and integrate it in time using a variation of the Euler method.
We use periodic boundary conditions in both space dimensions
which are practically enforced by the use of Fourier modes.
We typically use $128\!\times\! 128$ 
modes in the two space directions while our physical space has dimensions
$64\!\times\! 64$ units.
Due to the special treatment of the linear part, pseudospectral algorithms
allow for a relatively large time step. We typically take 
$\delta t\!=\!0.05$ when we use the parameter values
(\ref{eq:paramvalues}). 

\begin{figure}
  \epsfig{file=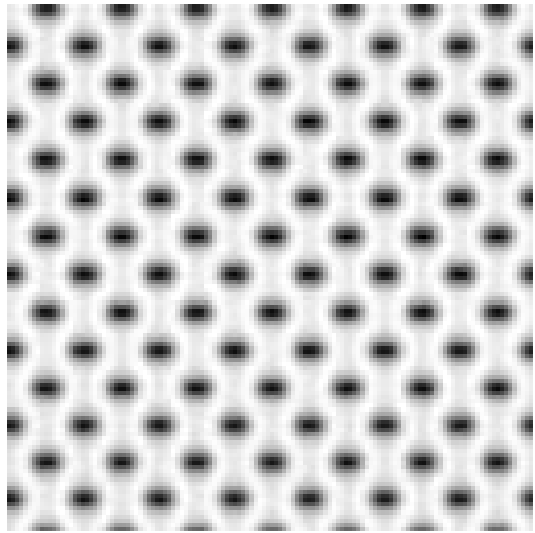,width=7.0cm}
  \caption{A two-dimensional periodically modulated pattern without
defects from a simulation of Eqs.~(\ref{eq:A2D.scaled},\ref{eq:phi2D.scaled})
for the parameter values (\ref{eq:paramvalues}), $c_1\!=\!c_2\!=\! 1$,
and $h\!=\!0.28$. Initial conditions: a slightly perturbed periodic modulation
state with period 9.14 .
Shown is a grey-scale representation of the field $|A|$. White corresponds
to the maximum value ($|A|\!=\!0.85$) and black corresponds to
the minimum ($|A|\!=\!0.12$).
The physical dimensions of the space are $64\!\times\!64$ and
we have used periodic boundary conditions.
 }
 \label{fig:2dpatternA}
\end{figure}

\begin{figure}
  \epsfig{file=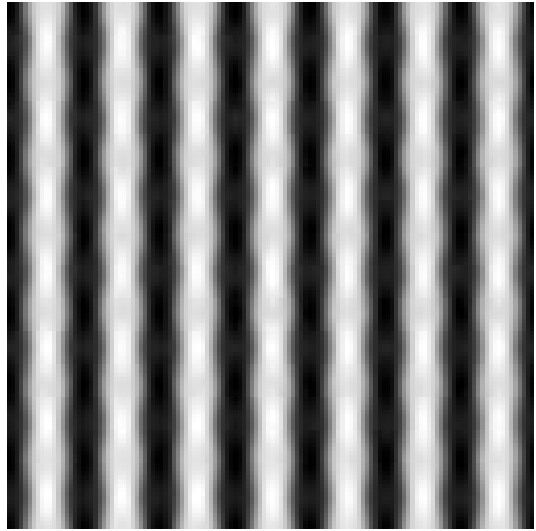,width=7.0cm}
  \caption{The field $\phi$ for the state of Fig.~\ref{fig:2dpatternA}.
White corresponds
to the maximum value and black corresponds to the minimum value
($\phi_m\!=\! \pm 0.998$).
  }
\label{fig:2dpatternphi}
\end{figure}

We start a simulation of Eqs.~(\ref{eq:A2D.scaled},\ref{eq:phi2D.scaled})
with a state that is periodic in $x$, e.g. with period 9.14.
The state is stable in the parameter range given in
Eq.~(\ref{eq:pstability}).
When we reduce the field $h$ slightly below the value 0.30 we find 
that the pattern is destabilized and modulations along the $y$
direction appear.  The pattern becomes periodic in both spatial
dimensions.
We give an example of such a state
in Figs.~\ref{fig:2dpatternA}, \ref{fig:2dpatternphi} where the
fields $|A|$ and $\phi$ are shown, respectively.
The state describes modulations in 2D and has no defects.
It was obtained for $h\!=\!0.28$.
We note that the state in Figs.~\ref{fig:2dpatternA},\ref{fig:2dpatternphi} 
is not fully static. It persist for a long time but it is eventually modified
through the creation of defects. The system presents persistent dynamics
until the end of our simulation, nevertheless, the 2D correlations
are preserved to a large extent.

In general, the 2D modulated states reached slightly below the
destabilization of 1D periodic 
modulations are delicate and we have not been able to find a truly static one. 
The near-periodicity of the final states
in both spatial directions is readily seen in the figures
resulting from our simulations.

\begin{figure}
   \epsfig{file=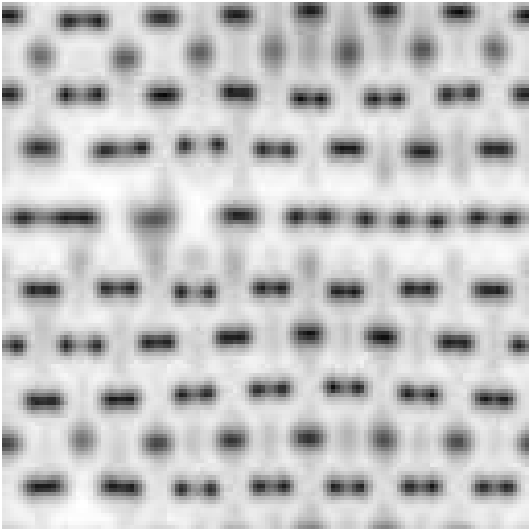,width=7.0cm}
   \caption{A reasonably developed defect lattice from simulations of
Eqs.~(\ref{eq:A2D.scaled},\ref{eq:phi2D.scaled})
for the parameter values (\ref{eq:paramvalues}),
$c_1\!=\!c_2\!=\! 1$, and $h\!=\!0.27$.
Initial conditions: a slightly perturbed periodic modulation state
with period 9.14. Shown is a grey-scale representation of $|A|$.
White corresponds to the maximum value and black corresponds to the minimum
($|A|\!=\!0$).
The state is dynamic but always remains close to the one shown.
The physical dimensions of our space are $64\!\times\! 64$ and we have used
periodic boundary conditions.
}
\label{fig:latticeA}
\end{figure}

\begin{figure}
   \epsfig{file=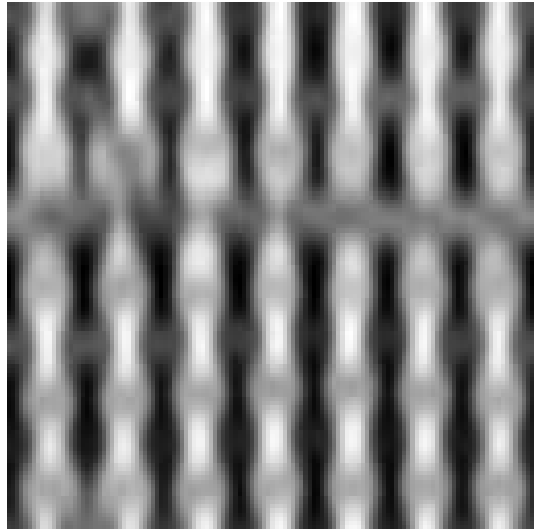,width=7.0cm}
   \caption{The field $\phi$ for the state of Fig.~\ref{fig:latticeA}.
White corresponds to the maximum value
and black corresponds to the minimum value.
}
   \label{fig:latticephi}
\end{figure}

An interesting point is that the defects which are typically
created in these processes also show strong 2D correlations.
In fact a slightly disordered defect lattice is the usual outcome of 
such numerical simulations.
An example of a reasonably developed defect lattice is
presented in Fig.~\ref{fig:latticeA} through the field $|A|$.
It was obtained by starting the simulation with a periodic state
with period 9.14. We used the value $h\!=\!0.27$ and the other
parameters as in (\ref{eq:paramvalues}). The initial 1D periodic state is then
unstable and evolves into the defect lattice.
The time evolution shows that the state has some substantial
dynamics but it always remains close to the picture of
Fig.~\ref{fig:latticeA}.
Finally, in Fig.~\ref{fig:latticephi} we give
the $\phi$ field of the defect lattice of Fig.~\ref{fig:latticeA}.
The $\phi$ field is predominantly varying in the $x$ direction
with small modulations in $y$.

\begin{figure}
   \epsfig{file=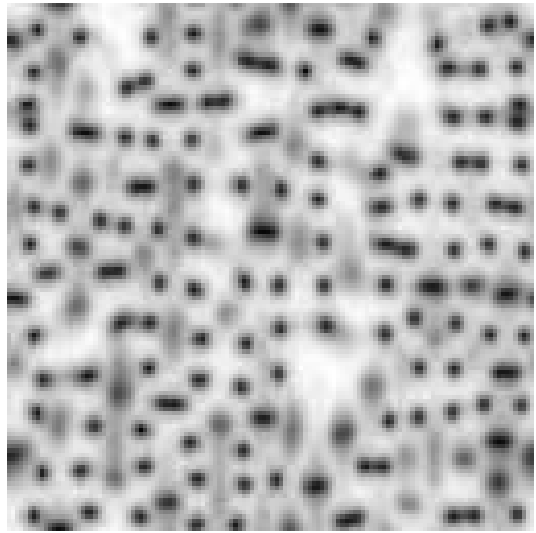,width=7.0cm}
   \caption{A disordered state: grey-scale representation of the field $|A|$.
The parameter values used are (\ref{eq:paramvalues}), $c_1\!=\!c_2\!=\! 1$
and $h\!=\!0.1$.
The space dimension are $64\!\times\! 64$ and we have used
periodic boundary conditions.
}
   \label{fig:chaotic}
\end{figure}

On further reducing the parameter $h$ the resulting pattern becomes
progressively more disordered. A defect chaotic state is eventually obtained
for small values of the parameter.
In Fig.~\ref{fig:chaotic} we show such a state for $h\!=\!0.1$.
These defect-chaotic states deserve to be studied on their own right
but a full investigation of this problem is beyond the scopes
of the present paper.

It is not easy to predict in what way the states that we study here would 
appear in an experiment. Details of the experimental procedure could be
important and the final result can be complicated as is
indicated, e.g., by Fig.~7 in \cite{BuTo00} where roughly periodic
states without defects are found.
In order to observe defect lattices in homeotropic cells one should remain 
at small values of $\varepsilon$ so that the director angle does not
become very large. Otherwise a transition to CRAZY rolls occurs first,
which entails formation of disclinations in the director field
\cite{RoEb00,BuTo00}. Also, at higher values of $\varepsilon$, 
mean-flow effects not included in our simple description become important.

Let us describe the protocol of a numerical simulation which may give
some guidance for experiments. The simulations are performed on a
$64\!\times\! 64$ domain. 
We start at a field $h$  above $h_{\rm AR}$ ($h\!=\!0.7$) 
with small spatially random initial conditions of the fields $A, \phi$.
Experimentally this corresponds to jumping from below the EC threshold
directly into the region of stable ARs.
Typically, large domains with both AR states appear, separated by walls
in the $x$ direction.
We then jump to $h\!=\!0.4$ where
ARs are destabilized while defects are created and a state
with 2D correlations sets in. The final state is roughly periodic
with period 12.8 in the $x$ direction (5 periods in the domain)
and 16 in the $y$ direction (4 periods in the domain).
The defects are also roughly ordered in a lattice.
After this we increase $h$ gradually in steps of $\delta h\!=\!0.02$ and
let the system relax in every step for 1000 time units.
Although the system never relaxes to a static state, we observe 
clearly correlations in both spatial directions with the defects
approximately ordered in arrays until $h\!=\!0.44$ is reached.
For $h\!\geq\!0.46$ the defects annihilate and the
system relaxes to a static 1D periodic state with period 12.8.
The process described above seems robust in our simulations.
Note the interesting fact that there is a range of stable
coexistence of the fully ordered
1D  state (it is stable down to $h=0.42$) and the 2D solution.

Defect lattices have in fact been observed in planar EC \cite{nasuno,BeJo97}
at fairly high frequencies. A theoretical description for these
systems by a much more 
elaborate quantitative theory has been developed recently \cite{berndthesis}.

%%%%%%%%%%%%%%%%%%%%%%%%%%%%%%%%%%%%%%%%%%%%%%%%%%%%%
\section{Static chevrons}    \label{chevrons}

Here we will relax the condition $c_1=c_2=1$ by reducing $c_1$
(slightly) below $1$.
This has no influence on the 1D solutions and their stability with respect to
$x$ dependent  fluctuations and increases the range of stability 
with respect to $y$ variations. In fact the critical value of $h$
for the zig-zag destabilization of the periodic solutions 
(dashed line in Fig.~\ref{fig:limits}) is reduced. 
For example, the period 9.14 is stable in the range
	\begin{equation}  \label{eq:c1c2pstability}
 	0.19 < h < 0.74\,, \quad \hbox{when}\; c_1=0.9,\; c_2=1\,,
	\end{equation}
which should be compared to Eq.~(\ref{eq:pstability}).
For values of $h$ beyond the instability the states that we
obtain are similar to the ones  found above for $c_1=1$.  
Namely, we observe states with 2D order close to the instability
point (with some hysteresis), which get progressively less ordered 
as $h$ is decreased. When $c_1$ is decreased further
the stable range of the periodic solutions extends to progressively lower
values of $h$. In fact, for $c_1<0.77$ the 1D solutions with period 9.14
are stable down to $h=0$.

\begin{figure}
   \epsfig{file=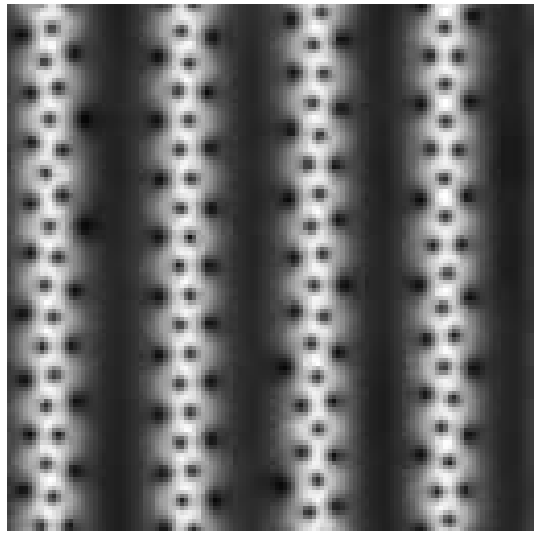,width=7.0cm}
   \caption{Static chevron state: grey-scale representation of the field $|A|$.
The parameter values used are (\ref{eq:paramvalues}),
$c_1\!=\! 0.9, c_2 \!=\! 1$, and $h\!=\!0$.
The physical dimensions is $64\!\times\! 64$ and we have used
periodic boundary conditions.
}
   \label{fig:chevronA}
\end{figure}

\begin{figure}
   \epsfig{file=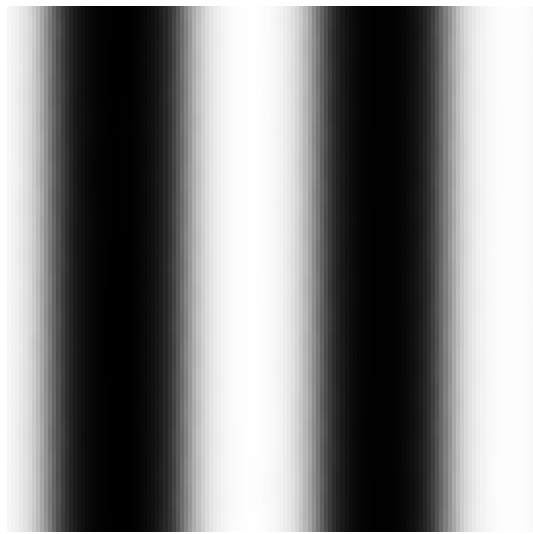,width=7.0cm}
   \caption{The field $\phi$ for the static chevron
state of Fig.~\ref{fig:chevronA}.
}
   \label{fig:chevronphi}
\end{figure}

For $h=0$ we obtain a novel static pattern containing lines
(or bands, i.e., multiple lines)
of defects along the $y$ direction which resemble the usual chevron states,
except that those are dynamic. 
In Figs.~\ref{fig:chevronA} and \ref{fig:chevronphi} such a solution
is presented for
$c_1=0.9$. The defects (zeros of $|A|$) are located at the dark points
in Fig.~\ref{fig:chevronA}. Within one band all defects have the same
topological charge and this is alternating from band to band.
The dark regions between defect bands imply a small value of $|A|$.
As one sees from Fig.~\ref{fig:chevronphi} the field $\phi$ varies
periodically essentially only in $x$, and thus is very reminiscent
of the 1D patterns discussed before.
We call these new patterns ``static chevrons''.
They are formed from random initial conditions. 

The static chevrons persist for small nonzero values of the parameter $h$. 
If we let the chevron states be created for $h\!=\! 0$ and then
increase $h$ the static chevrons persist up to $h\!\leq\!0.015$
for the parameters used here.
Increasing $h$ further ($h\!=\!0.02$) we observe 
that more defect bands are created, that is, we have chevrons with
a shorter period, but these are now the dynamic chevrons
presented previously for $c_1=1$ \cite{RoKra98,axelthesis}. 
For even larger values of $h$ one reaches the much less correlated defect
chaotic state discussed before.
For $c_1$ closer to $1$ the correlations become weaker.

To gain an understanding of static chevrons we first note that the
states between the
defect bands should be interpreted as (approximately) homogeneous states with 
nonzero wavenumber $P$ in the $y$ direction, i.e., with complex amplitude
	\begin{equation}  \label{eq:iPy}
	A = A_0\, e^{\pm i P y}\,,
	\end{equation}
where the sign in the exponent alternates between neighboring regions
and the magnitude of $P$ is related to the number of defects $\rho$
per unit length in one band through the relation
	\begin{equation}  \label{eq:defdensity}
	\rho = \frac{P}{\pi}\,.
	\end{equation}
This follows directly from the fact that each defect contributes
a phase change of $\pm 2 \pi$, depending on its topological charge. 
In the simulations we find that $\phi$ and $P$ take values close to 
	\begin{equation}  \label{eq:betweenchevrons}
	\phi_A=\pm\frac{1}{\sqrt{1-c_1^2}}\;\; \hbox{and}\; P_A=c_1 \phi_A\,,
	\end{equation}
in the region between the defect bands
while $A_0$ is small as was already mentioned.

\begin{figure}
   \epsfig{file=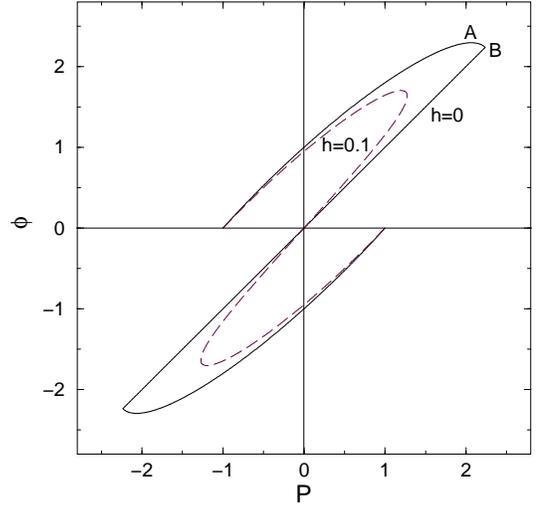,width=7.0cm}
   \caption{The solutions of Eq.~(\ref{eq:3rdorder}) (with $A_0^2\!>\!0$)
in the ($P - \phi_0$) plane for two values of the
field $h\!=\!0$ (solid line) and $h\!=\! 0.1$ (dashed line).
The parameter
values are $c_1\!=\! 0.9, c_2 \!=\! 1$. The point A has coordinates
$\phi_A\!=\!1/\sqrt{1-c_1^2},\,\, P_A\!=\! c_1 \phi_A$ and the point B has
$\phi_B\!=\! c_2/\sqrt{1+c_2^2-2 c_1 c_2},\, P_B\!=\!\phi_B/c_2$.
}
   \label{fig:phiP}
\end{figure}

The uniform solutions are described by equating the right-hand sides of
Eqs.~(\ref{eq:0D}) to zero. The resulting cubic equation for $\phi_0$
can be written as 
\begin{eqnarray}  \label{eq:3rdorder}
	 (\phi_0 - c_2 P)^3 - 2 \phi_0 (1-c_1 c_2) (\phi_0 - c_2 P)^2 
	\qquad \qquad \quad \nonumber \\
 	+ [(1+c_2^2-2 c_1 c_2) \phi_0^2 - c_2^2] (\phi_0- c_2 P) 
   	 + h\, c_2^2\, \phi_0 = 0 \,.
\end{eqnarray}
In the limit $h \to 0$ one of the solutions tends to
	\begin{equation}  \label{eq:phic2P}
	A_0^2 = 1 - \frac{1+c_2^2-2 c_1 c_2}{c_2^2} \phi_0^2\,, \quad
	\phi_0 = P c_2\,, 
	\end{equation}
and the other two tend to
	\begin{equation}  \label{eq:ARP}
	A_0=0\,, \quad \phi_0= c_1 P \pm \sqrt{1 - P^2 (1-c_1^2)}\,.
	\end{equation}
Here one has to impose the additional restriction that
$\phi_0/(\phi_0-c_2 P)>0$, 
since otherwise $A_0^2$ is not positive for $h$ slightly away from zero.
In Fig.~\ref{fig:phiP} $\phi$ is sketched for these limiting
solutions as a function of $P$ (solid lines). 
The solution (\ref{eq:phic2P}) is the continuation of the NRs to nonzero $P$.
Similarly, (\ref{eq:ARP}) represents the continuation of ARs.
The two lines join at point B in Fig.~\ref{fig:phiP}.

We can now see that the states between the defect bands
given in Eq.~(\ref{eq:betweenchevrons}) correspond (approximately) to the
continued ARs of Eq.~(\ref{eq:ARP}) with the {\it maximal} $|\phi_0|$.
Thus the defect bands connect between the two symmetry degenerate versions
of this state.  
A simple stability analysis explains why this state is selected. 
We write $A=A_0+a$ and $\phi=\phi_0+\varphi$ and linearize
Eqs.~(\ref{eq:A2D.scaled},\ref{eq:phi2D.scaled}) in $a,\ \varphi$. 
For $A_0=0$ the two equations separate and we obtain 
the following expression for the growth rate of the mode involving $a$
	\begin{equation} \label{eq:sigma}
	\tau \sigma =1-s_x^2-s_y^2+2 c_1 c_2 \phi_0 s_y - c_2^2 \phi_0^2\,.
	\end{equation}
One easily checks that  all states are unstable against fluctuations
along $y$ except for
the state (\ref{eq:betweenchevrons}), which is marginally stable.
The latter is denoted by the letter A in Fig.~\ref{fig:phiP}.

The static chevron states which are periodic and describe spatial oscillations
between the states (\ref{eq:betweenchevrons})
need not be themselves marginally stable. On the
contrary, they are numerically found to be
quite robust. This observation is in agreement
with the results of Sec.~\ref{stability} that the 1D periodic states are
stable even for parameter values for which the uniform states are unstable.

The effect of the value of $c_1$ on the chevron states is rather profound. 
This is seen by the (approximate) relation giving the
density of defects in the chevrons which can be found using
Eqs.~(\ref{eq:defdensity}) and (\ref{eq:betweenchevrons}):
	\begin{equation}  \label{eq:defdensity2}
	\rho = \frac{c_1}{\pi \sqrt{1-c_1^2}}\,.
	\end{equation}
Thus, as $c_1$ approaches $1$ the defect bands become broader (the separation 
between defects inside a band is rather independent of parameters). 
In the limit $c_1 \to 1$ one expects
existence of an infinite lattice of defects with the same polarity,
but this is not accessible numerically, and presumably also not experimentally.
On the other hand, as $c_1$ decreases, the number of defects per unit
length decreases quickly reaching a situation with a single chain in a band.
Decreasing $c_1$ further one reaches a critical value below which 
no defects are created. Instead one has the 1D periodic
modulations without defects which are here stable also for
$h\!=\!0$ (see above). 

We have mentioned already that static chevrons persist to slightly nonzero $h$.
In order to understand the defect-free regions for this case we have
studied the homogeneous solutions as obtained from Eq.~(\ref{eq:ARP}). 
In Fig.~\ref{fig:phiP} we also show $\phi_0$ for the case
$h\!=\! 0.1$ (dashed). Analyzing the stability we found that there is a
small interval around the states with maximal $|\phi_0|$ which are  stable.
This shows that the disappearance of static chevrons for increasing $h$
is not a result of the homogeneous regions becoming unstable,
but rather the defect bands
destabilize, which is also observed in simulations.

%%%%%%%%%%%%%%%%%%%%%%%%%%%%%%%
%%%%%%%%%%%%%%%%%%%%%%%%%%%%%%%
\section{Concluding remarks}    \label{conclusions}

Motivated in particular by experiments in electroconvection in
homeotropically aligned nematics \cite{RoEb00,BuTo00} we have studied some
classes of modulated solutions of the Ginzburg-Landau equation for the
complex order parameter $A$ describing the
bifurcation to a stationary roll pattern, coupled to a weakly damped
(or even undamped) homogeneous mode $\phi$ describing the orientation
of the in-plane director, see
Eqs.~(\ref{eq:A2D.unscaled},\ref{eq:phi2D.unscaled})  (unscaled) or 
Eqs.~(\ref{eq:A2D.scaled},\ref{eq:phi2D.scaled}) (scaled). 
The most important parameter $h$ in these equations gives the ratio of the
aligning torque on the director over the destabilizing torque of normal rolls.
The latter is proportional to the supercriticality parameter $\varepsilon$.
Another important parameter $\nu$ in the equations
characterizes the action of the gradient of the in-plane director
on the phase of the rolls, which can be controlled experimentally by
varying the frequency.
Our study is relevant for $\nu>0$, which is the case in the upper-frequency
conduction range and presumably also in the dielectric range \cite{Ro00}.

The solutions we considered are characterized by modulations in the 
direction perpendicular to normal rolls (parallel to the wave vector).
There exists a surprisingly rich spectrum of stable solutions of this
type even in the range of control parameter  where most homogeneous
states have lost stability. The simplest type has only variations in the 
direction of the wavevector (1D structures). 
When such variations arise they lead to the creation of defects,
except in a small 
control parameter region where 2D defect-free modulations may exist metastably.
The solutions with defects range from essentially fully spatially
ordered defect lattices 
to defect chaos with various degrees of spatial correlations. 
Their detailed study appears interesting, but is beyond the scope of this paper.
The complexity of the solutions, as well as their dynamics, generally
increase with decreasing $h$. For very small values of $h$, however,
defects show the tendency to 
order along chains (or bands) yielding chevron structures.
One mechanism for the generation of  dynamic chevrons has been related
to a Turing instability  \cite{RoKra98,axelthesis}.
We have found for $c_1<1$ a new type of fully ordered, static
chevrons that appear more related to the 1D structures.

Focusing on the in-plane (= ${\hat c}$) director field, the modulations
occur along the direction of prealignment of the  in-plane director
(from this point of view they give rise to ``bend'' distortions). 
In fact, there is a long history of observations of such modulated states
under an electric field in highly-doped MBBA
\cite{TruBliBa80,NaLuGi81,HuHiRo00,HuHiYu01} and other nematic
materials \cite{WePe88,EbToBu00} at high frequency, often without detectable
convection rolls. They come under the name of ``wide domains''
\cite{TruBliBa80,NaLuGi81,WePe88} or ``prechevrons'' \cite{HuHiRo00,HuHiYu01}.
Since such structures cannot be explained purely statically, it seems
reasonable to assume
that they are secondary structures of the type discussed here. 
The mechanism for the generation of the primary roll pattern,
which may not be visible, is presumably different from the usual
Carr-Helfrich mechanism. In fact, it was shown that the driving mechanism
persisted above the nematic-isotropic transition  \cite{HuHiYu01}.
Moreover, in \cite{WePe88} it was shown that a treatment of the bounding
plates by tensides led to a considerable increase of the frequency range
where the wide domains appeared.
These findings, which are consistent with some old results \cite{RiDu79},
indicate that an isotropic mechanism with rolls confined to the boundaries
is involved.

It was found that even after the wide domains have formed the usual
electroconvection would set in at the expected threshold
\cite{HuHiRo00,HuHiYu01}. 
In the dielectric regime this would lead directly to perfectly ordered
chevrons that have an appearance like the static chevrons presented above.

While states modulated only along the direction of the wavevector
have been observed to some extent, the observation of corresponding
two-dimensional modulations without defects is lacking.
This may be due to the small parameter range where such 2D
modulations occur and also because of their apparent metastability.
An experimental effort in this direction appears interesting.
The present results suggest that the objective should be to observe
successively the 1D and 2D modulations and the chevron states
and to test if they can be related to each other according to
the present theory.

Equations~(\ref{eq:A2D.scaled},\ref{eq:phi2D.scaled}) also allow for
solutions which are modulated only along the rolls (perpendicular to
the prealignment of the in-plane director, giving rise to a
``splay'' distortion). Then the phase of the patterning mode 
(i.e., the phase of $A$) comes into play, and the
periodic solutions in fact correspond to zig-zag patterns. Such states are
of relevance for smaller values of $\nu$, in particular for $\nu<0$.
Near the zig-zag instability of normal rolls, which occurs at $h=1-\nu$,
and thus precedes the transition to abnormal rolls for $\nu<0$, the
periodic patterns are unstable. 
For smaller  values of $h$ (larger $\varepsilon$) there are various
types of periodic patterns. The scenario is enriched by the fact that
the usual defects (point dislocations) can extend 
in the direction perpendicular to the rolls to form phase slip lines
\cite{hongthesis}.

In order to extend the treatment to larger $\varepsilon$ one has to
use more complicated equations derivable from an extended nonlinear analysis,
which includes in particular
mean flow. For planar convection in the conduction regime the analysis
has been carried out and relevant solutions involving in particular
realistic defect lattices have been studied \cite{berndthesis}. 
For the dielectric regime an extended weakly nonlinear study was presented
in \cite{Ro00}.
The system appears interesting in view of its parameters $c_1=c_2=\nu=0.98$
calculated under neglect of flexoelectric effects, which are of more
relevance in the dielectric regime (note that in Ref.~\cite{Ro00}
the $c_i$ correspond to our $C_i$).
There is hope to find static chevrons if the aligning torque could be
made sufficiently small by applying a destabilizing magnetic field.
One might use this technique also in the conduction range where
$c_1$ is substantially smaller than $1$ \cite{berndthesis}. 
However, one may then have to include higher-order terms in
Eq.~(\ref{eq:phi2D.scaled}).

\acknowledgments
We wish to thank \'A. Buka, N. \'Eber, B. Dressel, W. Pesch, and 
A.G. Rossberg for useful 
discussions. We are grateful to A.G. Rossberg for making available his
method and code for identifying defects \cite{defectlocate}.
Financial support from the European graduate school 
``Non-equilibrium phenomena and phase transitions in complex systems'',
by DFG under contract KR 690/16-1,
and from the TMR program ERBFMRXCT960085 funded by EU
is gratefully acknowledged. 
%%%%%%%%%%%%%%%

\end{document}